\documentclass[twocolumn,showpacs,amssymb,aps]{revtex4-1}
\usepackage{graphicx}
\usepackage{epsfig}
\usepackage{dcolumn}
\usepackage{bm}
\usepackage{color}

\def\lessim{\lower.5ex\hbox{$\; \buildrel < \over \sim \;$}}
\def\gtrsim{\lower.5ex\hbox{$\; \buildrel > \over \sim \;$}}
\newcommand{\ave}[1]{\left\langle #1 \right\rangle}

 \newcommand{\MeV}{\hbox{MeV}}

 \newcommand{\GeV}{\hbox{GeV}}

\begin{document} 
\topmargin -0.8cm\oddsidemargin = -0.7cm\evensidemargin = -0.7cm
\preprint{}

\title{Elliptic flow at varying energy heavy ion collisions: Partonic versus hadronic dynamics}

\date{March 2012}
\author{Vincenzo Greco$^{a,b}$, Michael Mitrovski$^{c}$, Giorgio Torrieri$^d$}

\affiliation{$\phantom{A}^a$Department of Physics and Astronomy, Via S. Sofia 64, I-95125 Catania, IT\\
$\phantom{A}^b$ INFN Laboratori Nazionali del Sud, Via S. Sofia 62, I-95125 Catania, IT \\
 $\phantom{A}^{c}$ Brookhaven National Laboratory, Upton New York 11973, USA \\ $\phantom{A}^d$ FIAS, Johann Wolfgang Goethe Universit\"at, \paragraph{}
Frankfurt}

\begin{abstract}
We examine whether the breakdown in elliptic flow quark number scaling observed at the Relativistic Heavy Ion Collider (RHIC) energy scan is related to the turning off of deconfinement by testing the hypothesis that hydrodynamics and parton coalescence  always apply, but are obscured, at lower energies, by variations in the widths of quark and anti-quark rapidity distribution.
We find that this effect is enough to spoil quark number scaling in elliptic flow. A lack of scaling in data therefore does  not signal the absence of partonic degrees of freedom and hadronization by coalescence.
In a coalescing partonic fluid, however, elliptic flow of anti-baryons should be greater than that of baryons, since anti-baryons contain a greater admixture of partons from the highly flowing mid-rapidity region. 
Intriguingly, purely hadronic dynamics has a similar dependence of baryon-anti-baryons  elliptic flow as purely partonic dynamics, again because anti-baryons tend to come from regions where the deviation of the system from hydrodynamic behavior is at its smallest.
The opposite trend observed in experiment is therefore an indication that we might be misunderstanding the origin of elliptic flow.  We finish by discussing possible explanations of this, and suggest experimental measurements capable of clarifying the situation.
\end{abstract}
\maketitle
\section{Introduction}
The discovery, at the relativistic heavy ion collider (RHIC), of a ``perfect fluid'' hadronizing via ``quark coalescence'' \cite{whitebrahms,whitephobos,whitestar,whitephenix,sqgpmiklos,heinzhydro,rev-coal} has aroused a lot of experimental and theoretical attention.   The main experimental evidence to support perfect fluidity is the observation that elliptic flow $v_2$, defined as the 2nd Fourier coefficient in the particle spectrum wrt the reaction plane $\phi_{rp}$
\begin{equation}
\frac{dN}{dy dp_T d\phi} = \frac{dN}{dy dp_T} \left( 1+ 2 v_n \cos\left( n(\phi-\phi_{rp}) \right)  \right)
\end{equation}
approaches the value expected from hydrodynamics \cite{heinzhydro}.
The further scaling of mesonic and baryonic $v_2$ is suggestive of what is seen by the most naive quark coalescence:
\begin{equation}
\label{coaldyn}
\left( \frac{dN}{d^3 p}  \right)_{hadron} = \int d^3 p_i \left(\prod_i \left( \frac{dN}{d^3 p}  \right)_{quark} \right)\times
\end{equation}
\[\ \times W\left(x_i|x_{hadron};p_i | p_{hadron}\right)\]
Provided the Wigner functions \cite{muellerpaper,grecopaper} $W(...)$ are $\delta$-functions in position and momentum, $v_2^{baryon} (p/3)/3=v_2^{meson}(p/2)/2$ \cite{molnarpaper} , a relation that holds surprisingly well at the top RHIC energy of 200 GeV per nucleon \cite{whitebrahms,whitephobos,whitestar,whitephenix,sqgpmiklos,heinzhydro}, even if signs of finite width Wigner functions
have been identified \cite{rev-coal,Greco:2007nu}.
The most common interpretation of this is that the degrees of freedom carrying elliptic flow are partonic \cite{rev-coal, muellerpaper,grecopaper,Greco:2007nu,hwapaper,molnarpaper,lacey}. Coalescence of partonic degrees of freedom is also hinted by the high $p_T$ baryon distribution \cite{muellerpaper,grecopaper}, and can account for 2-particle correlations structures such as the near-side ``ridge'' \cite{hwapaper} and the far-side ``cone'' \cite{machreco}. It has been found to significantly affect also the nuclear modification factor
and the elliptic flow of heavy quarks \cite{hq-coal1,hq-coal2}.

It is, however, far from clear how the flow data and its ``partonic ideal fluid'' interpretation  is related to the onset of deconfinement.   In the confined phase, the coupling constant between mesons goes as $\lambda/N_c^2$ \cite{thooft} while the density of degrees of freedom, impacting both viscosity $\eta$ and entropy density $s$, goes as $N_c^0$.   Hence, $\eta/s \sim N_c^2$.
In the deconfined phase the density of degrees of freedom goes as $N_c^2$, hence
$\eta/s \sim N_c^0$ independently of the coupling constant. $\eta/s$ should respectively go to a constant for strongly interacting theories \cite{viscmiklos} or go as $\sim ( \lambda^2 \ln \lambda)^{-1}$ in the perturbative regime \cite{amy}, both $\sim N_c^0$.   
Hence, Simple number of colors scaling \cite{thooft} shows that at deconfinement $\eta/s$ should jump by an order of magnitude.   If elliptic flow is indeed hydrodynamically generated, a corresponding jump should be observed when the initial temperature is about the critical temperature for deconfinement \cite{heinzscaling,mescaling1}. No such jumps in the elliptic flow observable are apparent in experimental data, whether varied in center of mass energy, centrality or rapidity \cite{whitephobos,mescaling1,mescaling2}.  

Seeing how the $v_2$ observable ``turns off'' is indeed one of the objectives of making {\em lower} energy measurements with the latest detectors \cite{low1,low2,low3,low4}.
Recently, the RHIC experimental energy scan started observing a systematic breakdown of coalescence in $\sqrt{s}$ (not seen, so far,in rapidity \cite{brahmsscaling})  which has been interpreted as signifying the ``turning off'' of partonic degrees of freedom \cite{scanv2,scanv2paper}.   However, as pointed out in \cite{soren}, such conclusion is premature:  even if quark coalescence persists, high baryochemical potential and small rapidity intervals in the collision region might be enough to make it not apparent when $v_2$ of mesons and baryons is considered.

In this work we explore the consequences of the rapidity dependence of both $dN/dy$ and $v_2(y)$ 
on the quark number scaling and on the baryon and anti-baryons splitting of the elliptic flow,
pointing out that recent observations of the RHIC experiments at low energy are not explicable
neither by a pure hydrodynamical+coalescence model nor by a pure hadronic model.
\begin{figure}[t]
\epsfig{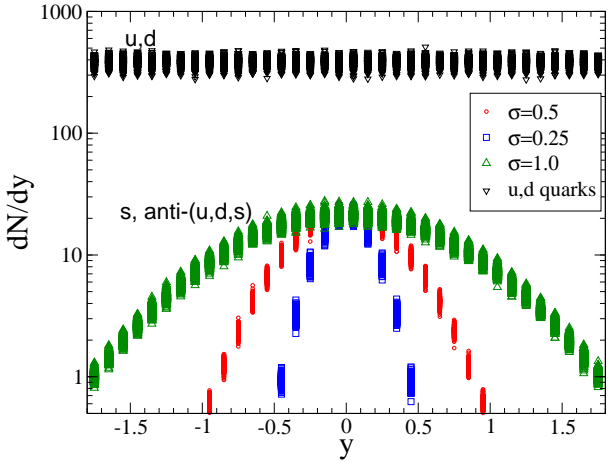}
\caption{\label{dndy} (color online) $dN/dy$ of quarks and anti-quarks. For anti-quarks we plot the 3
widths employed in our studies, $\sigma_y=0.25, 0.5,1$.}
\end{figure}
\section{MOdel}
Our model is simplified to capture the main issues we want to deal with.
We assume {\em all} $v_2$ is produced in a close-to-ideal hydrodynamic stage at {\em all} energies.   

The main consequence of an ideal hydrodynamics stage for our purposes is that, locally in rapidity and co-moving time, flow $u_\mu$ is species independent.   For configuration space coordinates $\vec{x}$, while {\em abundances} of flavors can vary,
\begin{equation}
\label{hydroass}
u_\mu^q(\vec{x}) = u_\mu^{\overline{q}} (\vec{x}) = u_\mu^s (\vec{x}) = u_\mu^{\overline{s}} (\vec{x})
\end{equation}
This is simply a consequence of the fact that in ideal fluids any conserved current $J^\mu \propto u^\mu$, and deviations from this relation are invariably dissipative \cite{denicol}.     This is {\em independent} of the initial condition for hydrodynamic evolution allowing to draw qualitative conclusions from an admittedly simplified schematic model.

Hence, before hadronization, provided flavor content is homogeneous in the transverse plane, for quarks emitted at rapidity $y$ and transverse momentum $p_T$ we have, up to mass corrections 
\begin{equation}
\label{v2eq}
 v_2^{q,s}(p_T,y) = v_2^{\overline{q},\overline{s}}(p_T,y)  ,\phantom{A}\ave{p_T}^{q,s} (y) = \ave{p_T}^{\overline{q},\overline{s}}(y)
\end{equation}

The other important ingredient to define the partonic stage is the $y$ and $p_T$ dependence of quark and antiquark
distribution function.
We assume a Gaussian antiquark distribution (Fig. \ref{dndy})
\begin{equation}
\label{dndyqbar}
\left(\frac{dN}{d^2 p_T\,dy}\right)_{\overline{q}} = N_{\overline{q}}^0 \exp \left[ -\frac{y^2}{\sigma_y^2} \right] f(p_T, \phi) 
\end{equation}
where $\sigma_y^2$ is approximately $\sim \ln s$, growing from $\simeq \ln (s/2m_{proton})$  to  $\simeq \ln (s/m_{proton})$ between the Landau and Bjorken limits \cite{landau,bjorken}.

For the quark distribution, we assume a flat distribution for a rapidity region $|y|<1$ meaning the probability of finding a valence quark is approximately invariant with rapidity. This appears quite reasonable 
at least if the beam energy is $\sqrt{s}> 5 \GeV$, 
\begin{equation}
\label{dndyq}
\left(\frac{dN}{d^2p_T\,dy} \right)_q = N_q^0 f(p_T,\phi)
\end{equation}
The normalization can be fixed by the $\overline{p}/p$ ratio
\begin{equation}
(dN_{\overline{q}}^0/dy)/(dN_q^0/dy) \sim \left[(dN_{\overline{p}}/dy)/(dN_p/dy)\right]^{1/3}
\end{equation}
experimentally decreasing with $\sqrt{s}$, as more entropy is carried by pre-existing valence quarks rather than created $\overline{q} q$ pairs.  
Strange quarks and antiquarks distribution is the same as antiquarks, since they are produced from zero in the hot medium. 

While these are undoubtedly simplified assumptions, more complicated scenarios \cite{jan} would, provided they model the experimentally observed limiting fragmentation of $v_2$ and $dN/dy$ \cite{mescaling1} and Eq. \ref{hydroass}, give qualitatively similar results to those we present.     This is because in any such system the bulk of antiquarks and strange quarks would come from a central plateau, whose width $\Delta y \leq 1$ grows very slowly with $\sqrt{s}$ \cite{whitephobos,mescaling1}.  The central plateau is also the region closest to ideal hydrodynamics, at all energies.

\begin{figure}[t]
\epsfig{width=8cm,figure=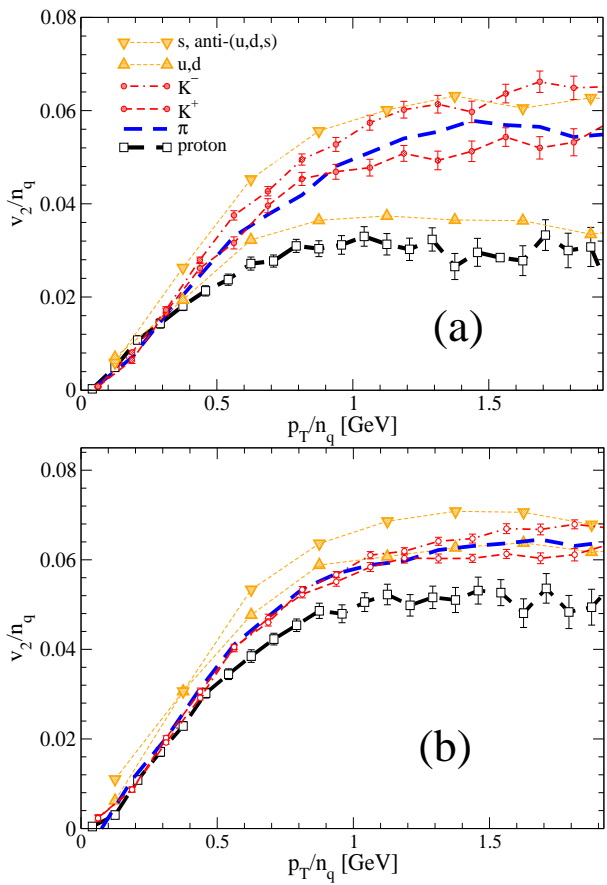}
\caption{\label{v2} (color online) $v_2/n_q$ vs $p_T/n_q$ of quarks, mesons baryons and anti-baryons in the $|y|\leq0.5$
rapidity window and for two different $\sigma$ values: Top panel (a) assumes $\sigma_y=0.25$ and bottom panel (b) has $\sigma_y=0.5$.}
\end{figure}

In accord with the hydrodynamic hypothesis for the quark phase, the momentum distribution is independent of the quark flavor ($q,\overline{q},s,\overline{s}$ have the same $\ave{p_T}$ and  $v_2$).  Hence, a similar $f(p_T,\phi)$ describes both $q$ and $\overline{q}$
\begin{equation}
\label{fp}
f(p_T, \phi) = e^{ -\frac{m_T}{T} } \left[ 1 + 2v_2^p(p_T,y)\cos \left(2 \phi \right) \right]
\end{equation}
and $T=170 \MeV$.
To fit experimental data \cite{mescaling1,mescaling2}, the quark (antiquark and strange quark) $v_2$ distribution for partons is also Gaussian: 
\begin{figure}[t]
\epsfig{width=8cm,figure=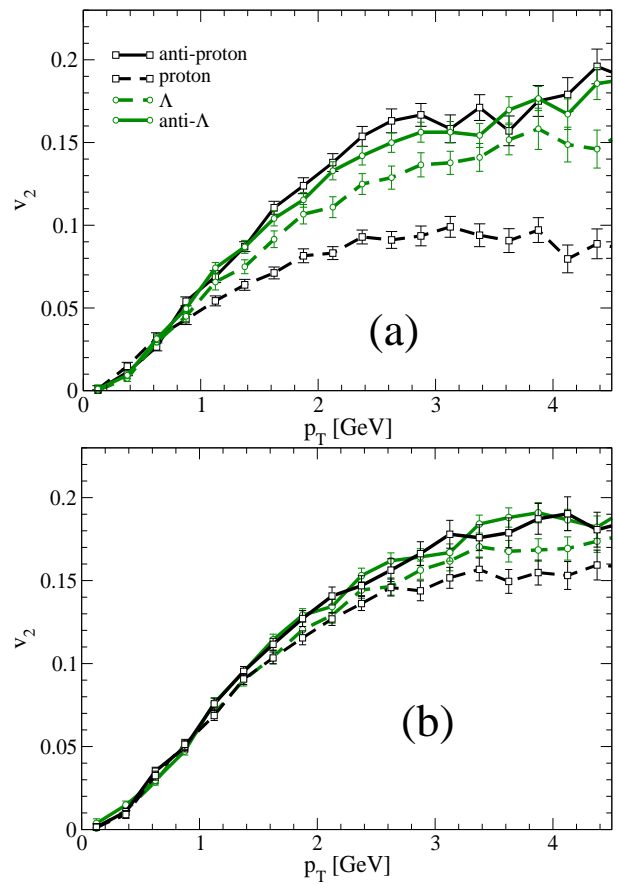}
\caption{\label{v2bar} (color online) $v_2$ of baryons and antibaryons calculated in the rapidity region $|y|\leq 0.5$.   The antiquark and strange quark distribution is a Gaussian with width in rapidity of, respectively, $\sigma_y= 0.25$ (top panel (a) ) and $\sigma_y=0.5$ (bottom panel (b) ). }
\end{figure}
\begin{equation}
v_2^p (p_T,y)  = v_2(y=0,p_T)  e^{ -\frac{y^2}{ \sigma_y^2} }
\end{equation}
$v_2^p(y=0,p_T)$ is an empirical function compatible with experimental data.
Its integral is fixed by the experimental constraint \cite{mescaling1} $v_2(y=0)/\epsilon \sim (1/S)(dN/dy)$, where S is the overlap area of the collision $S\simeq (1-\epsilon) \pi R^2$.  $v_2(y>0)$ is assumed, as seems to be the case in experimental data \cite{whitephobos,whitebrahms,brahmsscaling}, to have the same $\sigma_y$ as the ``hot'' medium (antiquark and strange quark distributions).
 
Note that even if $v_2(y,p_T)$ is the same for all partons {\em locally}, this is not true for $v_2(p_T)$ integrated over a finite $y$ region. For example, 
in Fig. 2, we show $v_2(p_T)|_{|y|\leq 0.5}$ for quarks (up triangles) and antiquarks/strange quarks (down triangles)  for two different widths. 

This difference is due to the different $dN/dy$ distributions for quarks and antiquarks together with the rapidity dependence of $v_2$. Anti-quarks have the largest $dN/dy$ from $y\sim 0$ region because at finite $y$ $dN/dy$ has a Gaussian-like tail
\begin{eqnarray}
v_{2, \overline q}(p_T)|_{|y| \leq y^*}=  \frac{{\int_{-y^*}^{ y^*}  {\! dy \, e^{ -\frac{y^2}{ \sigma_y^2}} v_2(p_T,y) }}}
{\int _{ -y^*}^{+y^*}  \! dy \,e^{ -\frac{y^2}{ \sigma_y^2}  }}
\end{eqnarray}
Since $v_2$ is maximum at $y \sim 0$, this generally leads to $v_{2,q}(p_T) < v_{2,\overline{q}}(p_T)$. 
Since the Wigner function is also approximately a Gaussian, a similar mechanism acts also in the coalescence process leading to a smaller flow for mesons
with both an $\overline{q}$ and an $s$ quark, such as $K^-$, with respect to those having a $q \overline{q}$ or $q \overline{s}$.

To study the $v_2$ of hadrons in the $p_T$ range $[0-5] \, \GeV$
we assume coalescence-type hadronization that for the case of mesons is given  by
\begin{eqnarray}
\label{eq:3a}
N_M = C_M\int \prod _{i=a,b} (p\cdot d\sigma)_i \,
{d^4{ p}_i} \,\delta(p_i^2-m_i^2) \nonumber\\
f_{i}(r_i,p_i)\, W\left( r|r_{hadron};q|q_{hadron} \right) \, .
\label{coal}
\end{eqnarray}
The relative phase space coordinates $r=r_b-r_a$ and
$q=p_b-p_a$ are the four-vector relative space-time distance and energy-momentum.
$d\sigma$ is a volume element of a space-like hyper-surface. The hyper-surface
of coalescing partons is fixed through the condition of equal 
longitudinal proper time $\tau=\sqrt{t^2-z^2}$ \cite{grecopaper,hq-coal1}.
 In the coalescence integral in Eq.(\ref{coal}) we consider the full phase space overlap of the coalescing 
particles, with the advantage of avoiding some of the more restrictive approximations 
employed by other coalescence models. In particular the extended 3D Wigner function avoids
any collinear approximations in momentum.

The hadron Wigner function for light quarks used is a simple 
product of spheres in position and momentum space
\[\
  W(r|r_{hadron};q|q_{hadron})=\frac{9\pi}{2} \Theta\left[\Delta_r^2-\left(r-r_{hadron}\right)^2\right]\times
\]
\begin{equation} \times \Theta\left[\Delta_p^2-\left( q-q_{hadron} \right)^2+\left(m_a-m_b\right)^2\right]. 
\end{equation}
The radii $\Delta_r$ and $\Delta_p$ in the Wigner formalism
obey the relation $\Delta_p =\Delta_r^{-1}$, motivated by the uncertainty
principle. Similar expression can be derived for 3-quarks coalescence. The parameter $\Delta_p $ is taken to be different for 
baryons and mesons and is of the order of the Fermi momentum, $\Delta^M_p=0.23 \, \rm GeV$ and $\Delta^B_p=0.35 \, \rm GeV$ .  
This has been shown to account for both the $\pi,K,p,\overline{p}, \Lambda$ spectra
at RHIC energy and 
quark number scaling elliptic flow including its small baryon/meson breaking \cite{Greco:2004ex,Greco:2004yc,Greco:2007nu}.
The condition for the 4-vector relative momentum can be written as:
\begin{eqnarray}
\label{prel}
( p_{T,a}- p_{T,b})^2-2m_{T,a}\,m_{T,b} \,  cosh(y_a-y_b)\nonumber\\
-m_{T,a}^2-m_{T,b}^2< \Delta^2_p
\end{eqnarray}
in the non-relativistic limit $|\vec p| <<m_i$ this reduces to the condition $(\vec p_a - \vec p_b)^2 \leq \Delta_p$, i.e.  
that the relative momentum, and hence also the difference in rapidity, of the coalescing quarks is limited to the width of the hadron wave function.

In the ultra-relativistic limit $|\vec p| >>m_i$ Eq.(\ref{prel}) reduces to:
\begin{equation}
2\, p_{T,a} \,p_{T,b}[cos(\phi_a-\phi_b)+cosh(y_a-y_b)]< \Delta_p^2
\end{equation}
which implies that at large $p_T$, due to the Lorentz boost from the fireball frame to the coalescing quark frame, particles with relatively large $\Delta p=|p_{T,a}-p_{T,b}|$ can coalesce.
This effect \cite{Greco:2007nu}  follows from relativistic kinematics and leads in our case to an increasing difference
between particle and anti-particle $v_2(p_T)$ as a function of $p_T$.

While in previous work the elliptic flow of coalescence was studied considering flat distribution in rapidity, 
we show that in presence of a finite Wigner function in Eq. \ref{coaldyn} and a $y-$dependent anti-quark density the naive quark number scaling breaks 
down and a difference in baryon-antibaryon $v_2$ appears {\em even if ideal partonic hydrodynamics and coalescence occur at all energies}.   While such a scenario is usually assumed not to apply for lower energies (this is very different from ``demonstratively falsified''), it is worthwhile to estimate the magnitude of these effects. 
\section{Results}
The meson and baryon $v_2$'s are summarized in Fig. \ref{v2}.   As can be seen, the
$v_n(p_T/n_q)/n_q$ naive scaling is considerably broken when $q-\overline{q}$ asymmetry is broken and the width of distribution is sufficiently narrow. When $\sigma^2_y=0.25$, the proton $v_2/n_q$ is below the meson one by nearly $50\%$ while the antiproton one is above it. $K^+$ and $K^-$ are similarly very different, while the difference between $\Lambda$ and $\overline{\Lambda}$ is smaller, see Fig. 3, because the difference comes
only from one over three quarks instead of one over two.
All of these can be readily explained by the greater admixture of high-flow antiquarks and strange quarks in anti-baryons and strange hadrons.  In the $K^-(s\overline u)$, for instance, {\em both} quarks tend to come from smaller $|y|$ where the flow is larger, while in $K^+$ only one does.  Hence, $K^+ (u\overline s)$ has a higher $v_2(p_T)$.
Up to an asymmetry of $\sigma_y=0.5$ the breaking of the scaling is still sizable but 
start to emerge when the transverse momentum per constituent quark is $\geq 1$ GeV. With $\sigma_y \ge 1$
the usual scaling are essentially restored.  
We notice that in a realistic description also thermal smearing should be included destroying the "locality" in momentum 
rapidity of the parton distribution and mixing quarks from flowing central region with the more baryon-rich fragmentation 
region. This effect with $\Delta y \simeq 0.5$, produces a similar effect to the one we have investigated for coalescence with
finite width wave function, therefore the lowering of baryon $v_2$  would persist even if $\Delta_{Wigner} \rightarrow 0$ as long as $\overline{p}/p$ breaks boost invariance.

\begin{figure}[t]
\epsfig{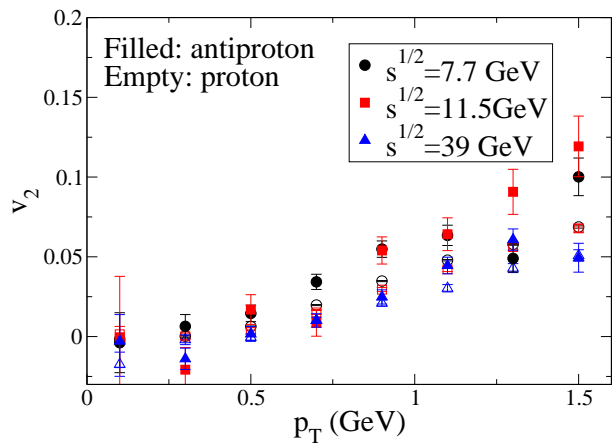}
\caption{\label{urqmd} (color online) $v_2$ of p and $\bar{p}$ at 7.7, 11.5 and 39 GeV for 0-80\% calculated within UrQMD}
\end{figure}

Our other main result, showing in Fig. \ref{v2bar}, is that $v_2$ is higher for anti-baryons than for baryons.   The difference is noticeable at all $p_T$ if the width in rapidity of the antiquark and strange quark distribution is $0.25$, but becomes noticeable at higher $p_T \sim 2$ GeV only, if the width increases to $\simeq 0.5$.   A further increase of the width makes the limiting $p_T$ of the difference even higher.

This result is stable against changes in the details of parton distributions as long as Eq. \ref{hydroass} and
the decreasing $\overline{q}/q$ as a function of rapidity are valid assumptions, since baryons admit an admixture of quarks from the high rapidity peripheral regions, where $v_2$ is smaller, while anti-baryons are dominated by central rapidity region at larger $v_2$ .  Any coalescence incorporating finite width wave function will result in a higher $v_2$ for anti-baryons w.r.t. baryons.
However, this directly contradicts experimental data \cite{scanv2}, suggesting either that coalescence breaks down at lower energies, or that quarks and antiquarks do not have the same flow $v_2(y,p_T)$.
\section{Discussion}
At this point one may ask if at lower energy what we have created is just a pure hadron gas.   The systematics of $v_2(p_T)$, overlapping for all accessible momenta between $\sqrt{s}$ of 7.7 GeV and LHC energies \cite{scanv2paper}, makes different regimes of $v_2$ origin immediately suspect.
Nevertheless, to investigate also this possibility we have  calculated the 
$v_2$ of protons and antiprotons in purely hadronic molecular dynamics, implemented via the UrQMD model \cite{urqmd}.

It should be noted that the version we used, v2.3 \cite{urqmd2.3}, suffers, as in earlier versions, from lack of detailed balance for multi-particle reactions: 
annihilation processes involving multi-particle final states, such as $p \overline{p} \rightarrow \pi \pi \pi \pi$ are possible.   The corresponding creation processes, however, are absent.  
Quantitatively, this is a minor correction, since its equilibration time for $2 \leftrightarrow n$ processes $\tau_{n \leftrightarrow 2} \sim e^{n-2} \tau_{2 \leftrightarrow 2}$, and, as uRQMD shows, even $2 \leftrightarrow 2$ processes do not equilibrate in realistic heavy ion collision expansion profiles.  However, in high chemical potential systems antiprotons themselves are also minor correction, being suppressed by factors of $\sim \exp(-(m+\mu_B)/T)$, so the {\em relative} importance of multi-particle processes could become enhanced.

The uRQMD result is shown in Fig. \ref{urqmd}, for the upper and lower physically relevant energies \cite{scanv2}, it has a {\em similar} behavior to purely partonic dynamics+coalescence.
   
The reason this happens is however somewhat different:
 Antiprotons are produced initially, and get absorbed by annihilation in a proton-rich medium.   
Neglecting regeneration, it is not surprising that in a hadronic medium protons and antiprotons propagate differently:  Their thermally averaged interaction cross-sections are considerably different, since protons typically interact via resonance formation while antiprotons, in a high chemical potential medium, can interact {\em both} via resonance formation and collisional processes and annihilation, with the latter dominating if the Knudsen number is large and $v_2$ is created by absorption rather than flow.   This, however, already makes it likely that antiprotons be {\em more} anisotropic, both due to flow and absorption, simply because they interact more with the medium.


Once again, naive extrapolation seems to disagree with 
experimental data \cite{scanv2},   {\em either} purely hadronic dynamics nor the coalescence hadronization of a partonic fluid.   
Ref.\cite{soren} proposed to explain this by assuming $v_2$ of particles transported from high rapidity is {\em higher} than the $v_2$ of particles created at mid-rapidity. It is however difficult to see how such an ansatz is compatible with the scaling of $v_2$ with rapidity \cite{whitephobos,mescaling1}, since  ``transported'' degrees of freedom have to come, and spend more time in, {\em less} flowing regions.    Our work confirms, in fact, that both partonic and hadronic dynamics, considering longitudinal diffusion {\em only}, have the opposite dependence on rapidity than that claimed in \cite{soren}.

As shown in \cite{steinh}, {\em transverse} ideal hydrodynamics, incorporating instant chemical and thermal equilibrium for all hadronic species, might provide a way out, via chemical composition anisotropy in the transverse plane.   
If flow is created close to the ideal hydrodynamic limit, outer regions in transverse space have significantly more flow than inner-lying regions.    The $\overline{p}/p \sim e^{-2 \mu_B/T}$ ratio, however, decreases with transverse radius since most $\overline{p}$ are formed in the higher density inner regions: in the outer regions $\mu_B$ is approximately the same but $T$ goes down, depleting the $\overline{p}/p$ ratio.
  
\begin{figure}[t]
\epsfig{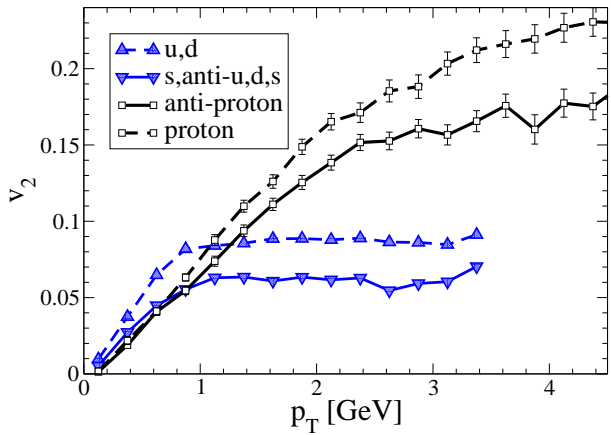}
\caption{\label{v2diff} (color online) Plot of baryon vs anti-baryons $v_2$ with same parameters as in Fig.\ref{v2bar} 
(upper panel) but with a modified width for quark, see text for details}
\end{figure}

As shown in \cite{steinh}, this effect can drive $v_2$ of protons above $v_2$ of antiprotons.      However, coalescence in this case is also not ruled out, since by exactly the same reasoning, given a partonic system with $q-\overline{q}$ transverse anisotropy, one can drive the mean flow of quarks above the mean flow of anti-quarks, effective relaxing the assumption of 
Eq. \ref{v2eq}, $v_{2q}(y)=v_{2\overline q}(y)$: {\em locally in transverse space} flow is of course still the same for all species, but the inhomogeneity leads to different transverse-averaged flows. Fig. \ref{v2diff} shows what happens if the effective $v_{2q}$ is just 20$\%$ of those of the antiquarks, a comparable effect to that used in \cite{steinh} (the exact magnitude of the $q-\overline{q}$ transverse anisotropy should be approximately insensitive to whether uRQMD or ``partonic'' initial conditions are used). 
Partonic chemical non-equilibrium at freeze-out might accentuate this imbalance \cite{dusling}, since at $\mu_B \gg T$ light quarks equilibrate radiatively while light antiquarks equilibrate collisionally.

We conclude, therefore, that 2+1 {\em partonic} hydro+coalescence might also be consistent with lower energies RHIC data if the transverse radius dependence of elliptic flow is taken into account.    
Since, as discussed previously, strange quarks are chemically similar to light antiquarks in that they are created thermally, we expect, if this scenario is correct, that\\
$v_2^{p}-v_2^{\overline{p}}
>  v_2^{\Lambda}-v_2^{\overline{\Lambda}} >  v_2^{\Xi}-v_2^{\overline{\Xi}} > v_2^{\Omega}-v_2^{\overline{\Omega}} \phantom{A},\phantom{A} v_2^{\Omega} \simeq v_2^{\overline{\Omega}} $.

Alternatively, differential quark-antiquark flow can be directly realized in a system with either partonic or hadronic degrees of freedom {\em and} strong vector mean fields, since the repulsive channel at finite net quark number will generally be greater for $q,s$ than $ \overline{q},\overline{s}$ (NB,both light and strange quarks).     Vector mean fields, unlike scalar ones, can produce a difference between quarks and antiquarks because they admit both attractive and repulsive channels, and are sensitive to a conserved charge density such as the net baryon or quark density.

In a medium with positive chemical potential, vector mean fields should cause relatively more attraction between anti-particles and the higher baryonic density regions, and therefore a smaller  $v_2(p_T)$ for antiparticles with respect to the particles. 
 As shown in \cite{plumari}  within Boltzmann transport theory an attractive mean field cause a significant reduction of the
$v_2(p_T)$. This can ultimately be related, in a hydrodynamical picture, to the fact that an attractive mean field decreases the effective
pressure of the system at equilibrium.  While elliptic flow is thought to be driven by pressure gradients,  it remains to be clarified up to what level the transport and hydrodynamic picture are equivalent.

At hadronic level the vector mean fields are, ultimately how the AMPT calculation in \cite{similar} has been able to reproduce both the baryon and anti-baryon $v_2$ data.

A model with {\em quark} mean fields has also been recently conjectured and studied within NJL \cite{plumari} and PNJL models \cite{Bratovic:2012qs}.
As \cite{koch,biro,phsd} have shown quark degrees of freedom coupled to mean fields can produce  sizable elliptic flows \cite{plumari,Song-vector} which, in a vector channel, can again vary considerably between quarks and antiquarks.

If, in the partonic system, the effect of mean fields is comparable to isotropic pressure, however, the very definition of thermalization and $\eta/s$ needs to be revised, since defining a co-moving frame within a strong mean field is impossible: Particles and antiparticles will always move in different directions within a volume element, exactly the effect required to explain energy scan data.
This also means that if $v_2$ is driven by mean fields, one would expect the difference between baryon and anti-baryon $v_2$ to be 
constant with strangeness, since QCD dynamics is  flavor-blind\\ $v_2^{p}-v_2^{\overline{p}}
\simeq  v_2^{\Lambda}-v_2^{\overline{\Lambda}} \simeq  v_2^{\Xi}-v_2^{\overline{\Xi}} \simeq  v_2^{\Omega}-v_2^{\overline{\Omega}}$\\
This latter scenario, however, leads to some unsettling questions for the interpretation of $v_2$ at {\em higher} energies:  Since at $\overline{p}/p \simeq 1$ such effects become invisible (C-symmetry of the medium is restored), there is no indication that they actually turn off at RHIC/LHC, and the smoothness of the decrease of the $v_2^p-v_2^{\overline{p}}$ with $\sqrt{s}$ \cite{scanv2} suggests they do not.    As mean fields are ``large'' deviations from the hydrodynamic limit, discovering that they have a contribution to {\em creating} (rather than suppressing) $v_n$ could significantly modify viscosity estimates such as those at higher energies \cite{heinzhydro}.

To constrain the models further using existing data, let us consider that
the {\em total} ($\pi,K,p$) $v_2(p_T)$ in the same energy scan can be experimentally shown \cite{scanv2paper} to be approximately independent of the initial density (which depends, approximately, as $\frac{1}{S}\frac{dN}{dy} \sim N_{part} \ln \sqrt{s}$ \cite{whitephobos} assuming Bjorken expansion \cite{bjorken} with a constant starting time);  The integrated $v_2$ varies because $\ave{p_T}$ increases systematically with $\frac{1}{S}\frac{dN}{dy}$, but $v_2(p_T)$ overlaps within error bar \cite{scanv2paper}.    

The analysis in Fig. \ref{urqmd} also fails to reproduce such a scaling specifically for protons, in line with the interpretation of absorption as the dominant origin of $v_2$;   In this case, the $v_2$ scaling with $p_T$ should be governed by the collision integral formula of \cite{heisel}, where transverse density enters both into $\ave{p_T}$ and $v_2(p_T)$ via
\begin{equation}
v_2(p_T) \sim  \ave{\sigma v_{ij}(p_T)} \left( \frac{\epsilon}{S}\frac{dN}{dy} \right) 
\end{equation}
Here $\ave{\sigma v_{ij}(p_T)}$ is the normalized cross-section times negative velocity for {\em that particular transverse momentum}.   Neglecting averaging over rapidity, and assuming $d\sigma_{ij} \sim \alpha_{ij} dp_{TJ}^2/Q^4$ where $Q$ is the momentum transfer and $p_{TJ}$ is the momentum of the second scatterer, we get
\begin{equation}
v_2(p_T) \sim \epsilon \frac{dN}{dy} f \left(\frac{p_T}{\ave{p_T}} \right)\frac{G(p_T)}{S}
\end{equation}
\[\  
G(p_T) =\alpha_{ij} \int_0^{2\pi} d\phi \int_0^\infty f\left( \frac{p_{TJ}}{\ave{p_{T}}} \right) \frac{p_{TJ} d p_{TJ}}{\left( p_T^2 + p_{TJ}^2 - 2 p_T p_{TJ} \cos(\phi) \right)^2} \]
where we used the universal scaling  of the transverse momentum distribution with $p_T/\ave{p_T}$ noted in \cite{hwascaling} and $f(...)$ is a distribution function, which can be well approximated by a Tsallis spectrum \cite{tsallis}.

  The $G(p_T)$ term generally breaks the scaling, increasing $v_2(p_T)$  by virtue of an increase in $\ave{p_T}$ with $\sqrt{s},N_{part}$ and decreasing $v_2(p_T)$ at the high $p_T \gg \ave{p_T}$ tail. 
in case of high baryochemical potential limit at low $\sqrt{s}$, the averaging
of $\ave{p_T}$  over $p,\pi,K,...$  and $\alpha_{ij}$ over $p-p,p-\pi,\pi-\pi,...$ further breaks the scaling.
This systematic shift is also visible in Fig.11 of \cite{scanv2paper} or in \cite{sorensennew} within the uRQMD and AMPT models, as $v_2(p_T)$ increases systematically with $\sqrt{s}$.
The same reasoning explains the systematic rise specific to $p,\overline{p}$ observed in Fig. \ref{urqmd}.  

  However, as explicitly discussed in \cite{proc,comingwork}  but clear from \cite{heinzcf,heinzhydro}, an ideal fluid stage creating $v_2$ via azimuthally inhomogeneus pressure gradients will {\em not} ameliorate this problem, since a Cooper-Frye freeze-out \cite{cf} will result in a $v_2$ which to leading order depends on the 2nd Fourier components of the freeze-out hyper-surface $\epsilon_2^{\Sigma_T} $ and transverse flow $ \epsilon_2^{ v_T}$ via
\begin{equation}
v_2(p_T) \sim \ave{   \tanh \left[  \frac{\gamma \epsilon_2^{v_T}   p_T }{T} \right] \left(E - p_T \epsilon_2^{\Sigma_T}   \right)+ p_T \epsilon_2^{\Sigma_T}  }
\end{equation}
and the averaging $\ave{...}$ is done over events at a given $\sqrt{s}$ and centrality class.

Since both $\epsilon_2^{\Sigma_T}$ and $\epsilon_2^{v_T}$ depend on the lifetime of the system (albeit approximately saturate after a finite time \cite{v2orig}), $v_2(p_T)$ should also systematically rise with $\sqrt{s}$. The systematic deviation should increase with $p_T$, since both $\epsilon_2^{\Sigma_T,v_T}$ come with factors $p_T$. Such a behavior is indeed evident in \cite{heinzhydro}, and is qualitatively the same as the transport simulation \cite{sorensennew}.

In the recent years it has been point out that an agreement with data for several observables even at the highest RHIC energy can be better explained if one introduce a core-corona model \cite{nantes1}. 
The fact that AMPT, which incorporates both hadronic and partonic components in the relevant energy density regimes, has the same qualitative dependence as \cite{heinzhydro} (as seen in \cite{scanv2paper} Fig. 11) indicates that a core-corona model might find it challenging to reproduce $v_2(p_T)$ for all energies, since the greater preponderance of the corona at lower $\sqrt{s}$ should also change the $p_T$ dependence of $v_2$ ($v_2$ in the corona is either zero or hadronic).   However, since a core-corona model with an ideal core ($\eta/s=0$) does a good job of reproducing integrated $v_2/\epsilon$ at several top energies and centralities \cite{nantes1}, and $\eta/s=0$ is expected to scale the best \cite{heinzscaling}, perhaps the scaling of $v_2$ will to a certain extent survive when binned in $p_T$. This can be expected also on the base of dynamical transport calculations that include the strong $\eta/s$ increase at hadronic energy densities \cite{Ferini:2008he}. 
Regarding the baryon/anti-baryon pattern of $v_2(p_T)$, we have shown in the present paper that both a partonic and a hadronic picture show the same baryon/ anti-baryon pattern. Therefore we can say that also employing a core-corona picture the issue raised might persist unless strong transverse gradients within the core, a la \cite{steinh}, are maintained. 
It is therefore difficult to see whether either a hybrid analysis of a uRQMD-only analysis can accommodate the difference between $p$ and $\overline{p}$ as well as the near-overlap of $v_2(p_T)$ seen in \cite{scanv2paper}.
This near-overlap has not, however, as yet been established for separate particle species, so perhaps baryonic $v_2$ is quantitatively different from $\pi$-dominated total $v_2$.  In this case, hydrodynamic evolution with a long hadronic afterburner \cite{steinh} might be the simplest explanation of elliptic flow systematics.

In conclusion, we used quark coalescence and a simple parametrization of a quark fireball with hydrodynamic flow to investigate the dependence of coalescence-type scaling at various energies.   We concluded that it is wrong to expect coalescence to give the naive quark number scaling of $v_2$ at all energies and system sizes {\em even if partonic hydrodynamics and coalescence actually occur}.  We have also found that both coalescence and hadronic dynamics predict the opposite  difference of baryon and anti-baryon $v_2$ w.r.t. experimental data.  The scaling of baryonic $v_2$, therefore, points to the existence of separate quark and antiquark flow  at lower energies. A theoretical explanation of this will be essential to link flow dynamics in heavy ion collisions to the deconfinement phase transition and this is
presently missing.
 
G.T. acknowledges the financial
support received from the Helmholtz International Center for FAIR within the framework of the LOEWE program (Landesoffensive zur Entwicklung
Wissenschaftlich-\"Okonomischer Exzellenz) launched by the State of Hesse.
V.G. acknowledges the support by the ERC-StG2010 under the QGPDyn Grant.
M.M. acknowledges the support by the Brookhaven Science Associates, LLC
under Contract No. DE-AC02-98CH1-8886  with the U.S. Department of Energy.
We would like to thank Jan Steinheimer-Froschauer for helpful discussions and sharing details of not yet published work with us.

\vskip 0.3cm



\begin{thebibliography}{19}
\providecommand{\bibinfo}[2]{#2}
\bibitem{whitebrahms}
 I.~Arsene {\it et al.}  [BRAHMS Collaboration],
perspective
  Nucl.\ Phys.\ A {\bf 757}, 1 (2005)

\bibitem{whitephobos}
  B.~B.~Back {\it et al.},
  Nucl.\ Phys.\ A {\bf 757}, 28 (2005)

\bibitem{whitestar}
  J.~Adams {\it et al.}  [STAR Collaboration],
quark  gluon
evidence from
  Nucl.\ Phys.\ A {\bf 757}, 102 (2005)
\bibitem{whitephenix}
 K.~Adcox {\it et al.}  [PHENIX Collaboration],
nucleus
collaboration,''
  Nucl.\ Phys.\ A {\bf 757}, 184 (2005)
  
\bibitem{sqgpmiklos}
  M.~Gyulassy and L.~McLerran,
  Nucl.\ Phys.\  A {\bf 750}, 30 (2005)

\bibitem{heinzhydro}
  H.~Song, S.~A.~Bass, U.~Heinz, T.~Hirano and C.~Shen,
  Phys.\ Rev.\ Lett.\  {\bf 106}, 192301 (2011)

\bibitem{rev-coal}
  R.~J.~Fries, V.~Greco and P.~Sorensen,
  Ann.\ Rev.\ Nucl.\ Part.\ Sci.\  {\bf 58} (2008) 177

\bibitem{muellerpaper}
  R.~J.~Fries, B.~Muller, C.~Nonaka and S.~A.~Bass,
  Phys.\ Rev.\  C {\bf 68}, 044902 (2003)


\bibitem{grecopaper}
  V.~Greco, C.~M.~Ko and P.~Levai,
  Phys.\ Rev.\  C {\bf 68}, 034904 (2003)


\bibitem{molnarpaper}
  D.~Molnar and S.~A.~Voloshin,
  Phys.\ Rev.\ Lett.\  {\bf 91}, 092301 (2003)


\bibitem{Greco:2007nu}
  V.~Greco,
  Eur.\ Phys.\ J.\ ST {\bf 155} (2008) 45
  
\bibitem{hwapaper}
  R.~C.~Hwa and C.~B.~Yang,
  Phys.\ Rev.\  C {\bf 66}, 025205 (2002)

\bibitem{lacey}
  A.~Adare {\it et al.}  [PHENIX Collaboration],
  Phys.\ Rev.\ Lett.\  {\bf 98}, 162301 (2007)

\bibitem{machreco}
  V.~Greco, G.~Torrieri, J.~Noronha and M.~Gyulassy,
  Nucl.\ Phys.\  A {\bf 830}, 785 (2009)

\bibitem{hq-coal1}  
  V.~Greco, C.~M.~Ko and R.~Rapp,
  Phys.\ Lett.\ B {\bf 595} (2004) 202
  [nucl-th/0312100].
  
\bibitem{hq-coal2}
  H.~van Hees, M.~Mannarelli, V.~Greco and R.~Rapp,
  Phys.\ Rev.\ Lett.\  {\bf 100} (2008) 192301

\bibitem{thooft}
  G.~'t Hooft,
  Nucl.\ Phys.\  B {\bf 72}, 461 (1974).

\bibitem{viscmiklos}
  P.~Danielewicz and M.~Gyulassy,
  Phys.\ Rev.\  D {\bf 31}, 53 (1985).

\bibitem{amy}
  P.~B.~Arnold, G.~D.~Moore and L.~G.~Yaffe,
  JHEP {\bf 0011}, 001 (2000)

\bibitem{heinzscaling}
  H.~Song and U.~W.~Heinz,
  Phys.\ Rev.\  C {\bf 78}, 024902 (2008)

\bibitem{mescaling1}
  G.~Torrieri,
  Phys.\ Rev.\  C {\bf 82}, 054906 (2010)

\bibitem{mescaling2}
  G.~Torrieri, 
  Phys.\ Rev.\  C {\bf 76}, 024903 (2007)

\bibitem{low1}
        M.~Posiadala [NA61 Collaboration],
        arXiv:0901.3332. 

\bibitem{low2}
        A.~N.~Sissakian, A.~S.~Sorin [NICA Collaboration],
        J.~Phys.~G {\bf 36}, 064069 (2009).

\bibitem{low3} 
  M.~M.~Aggarwal {\it et al.} [STAR Collaboration],
  arXiv:1007.2613 [nucl-ex].

\bibitem{low4}
        P.~Staszel [CBM Collaboration],
        Acta Phys.~Polon.~B {\bf 41}, 341 (2010).


\bibitem{brahmsscaling} 
  F.~Videbaek [BRAHMS Collaboration],
  Nucl.\ Phys.\ A {\bf 830}, 43C (2009)


\bibitem{scanv2} 
  S.~Shi [The STAR Collaboration],
  arXiv:1111.5385 [nucl-ex].

\bibitem{scanv2paper} 
  L.~Adamczyk {\it et al.}  [STAR Collaboration],
  arXiv:1206.5528 [nucl-ex].

\bibitem{soren} 
  J.~C.~Dunlop, M.~A.~Lisa and P.~Sorensen,
  Phys.\ Rev.\ C\ {\bf 84}, 044914  (2011)




\bibitem{denicol} 
  B.~Betz, G.~S.~Denicol, T.~Koide, E.~Molnar, H.~Niemi and D.~H.~Rischke,
  EPJ Web Conf.\  {\bf 13}, 07005 (2011)
G. Denicol, D.Rischke et al, in preparation

\bibitem{landau}
  C.~Y.~Wong,
  arXiv:0809.0517 [nucl-th].

\bibitem{bjorken}
  J.~D.~Bjorken,
  Phys.\ Rev.\  D {\bf 27}, 140 (1983).

\bibitem{jan} 
  J.~Letessier and J.~Rafelski,
  J.\ Phys.\ G G {\bf 28}, 183 (2002)
  [hep-ph/0106151].

\bibitem{Greco:2004ex} 
  V.~Greco and C.~M.~Ko,
  Phys.\ Rev.\ C {\bf 70}, 024901 (2004); 
  
 \bibitem{Greco:2004yc}
  V.~Greco, C.~M.~Ko and I.~Vitev,
  Phys.\ Rev.\ C {\bf 71} (2005) 041901
   
\bibitem{urqmd}
  M.~Bleicher {\it et al.},
  J.\ Phys.\ G {\bf 25}, 1859 (1999),
  S.~A.~Bass {\it et al.},
  Prog.\ Part.\ Nucl.\ Phys.\  {\bf 41}, 255 (1998),

\bibitem{urqmd2.3}
  H.~Petersen, M.~Bleicher, S.~A.~Bass and H.~Stocker,
  arXiv:0805.0567 [hep-ph].

\bibitem{steinh} J.Steinheimer-Froschauer, V.Koch,M. Bleicher, 
 arXiv:1207.2791 [nucl-th].\\
J.Steinheimer-Froschauer, private communication


\bibitem{dusling}
  K.~Dusling, G.~D.~Moore and D.~Teaney,
  Phys.\ Rev.\  C {\bf 81}, 034907 (2010)



\bibitem{plumari}
  S.~Plumari, V.~Baran, M.~Di Toro, G.~Ferini and V.~Greco,
  Phys.\ Lett.\  B {\bf 689}, 18 (2010)


\bibitem{similar} 
  J.~Xu, L.~-W.~Chen, C.~M.~Ko and Z.~-W.~Lin,
  Phys.\ Rev.\ C {\bf 85}, 041901 (2012)
  [arXiv:1201.3391 [nucl-th]].

\bibitem{Bratovic:2012qs}
  N.~M.~Bratovic, T.~Hatsuda and W.~Weise,
  arXiv:1204.3788 [hep-ph].



\bibitem{koch}
  V.~Koch,
  Nucl.\ Phys.\  A {\bf 830}, 479C (2009)



\bibitem{biro}
  T.~S.~Biro, M.~Gyulassy and Z.~Schram,
  arXiv:1111.4817 [hep-ph].



\bibitem{phsd}
  E.~L.~Bratkovskaya, W.~Cassing, V.~P.~Konchakovski, O.~Linnyk, V.~Ozvenchuk and V.~Voronyuk,
  J.\ Phys.\ Conf.\ Ser.\  {\bf 316}, 012027 (2011)



\bibitem{Song-vector}
V. Greco, C. M. Ko, S. Plumari, T. Song, in preparation.  

  
\bibitem{heisel} 
  H.~Heiselberg and A.~-M.~Levy,
  Phys.\ Rev.\ C {\bf 59}, 2716 (1999)
  [nucl-th/9812034].


\bibitem{hwascaling} 
  R.~C.~Hwa and C.~B.~Yang,
  Phys.\ Rev.\ C {\bf 67}, 064902 (2003)
  [nucl-th/0302006].

\bibitem{tsallis} 
  T.~S.~Biro and B.~Muller,
  Phys.\ Lett.\ B {\bf 578}, 78 (2004)
  [hep-ph/0309052].

\bibitem{sorensennew} 
  D.~Solanki, P.~Sorensen, S.~Basu, R.~Raniwala and T.~K.~Nayak,
  arXiv:1210.0512 [nucl-ex].


\bibitem{proc} 
  G.~Torrieri, B.~Betz and M.~Gyulassy,
  arXiv:1208.5996 [nucl-th].

\bibitem{comingwork} G.Torrieri,B.Betz,M.Gyulassy, to be published



\bibitem{heinzcf}
  E.~Schnedermann, J.~Sollfrank and U.~W.~Heinz,
  Phys.\ Rev.\  C {\bf 48}, 2462 (1993)



\bibitem{cf}
  F.~Cooper and G.~Frye,
  Phys.\ Rev.\  D {\bf 10}, 186 (1974).


\bibitem{v2orig}
  J.~Y.~Ollitrault,
  Phys.\ Rev.\ D {\bf 46}, 229 (1992).


\bibitem{nantes1} 
  J.~Aichelin and K.~Werner,
  J.\ Phys.\ G G {\bf 37}, 094006 (2010)


\bibitem{Ferini:2008he}
  G.~Ferini, M.~Colonna, M.~Di Toro and V.~Greco,
  Phys.\ Lett.\ B {\bf 670} (2009) 325 

\end{thebibliography}
\end{document}